\documentclass{ws-mplb}

\begin{document}

\markboth{F. Pennini, A. Plastino, M. C. Rocca}{Classical
thermodynamics from quasi-probabilities}

%%%%%%%%%%%%%%%%%%%%% Publisher's Area please ignore %%%%%%%%%%%%%%%
%
\catchline{}{}{}{}{}
%
%%%%%%%%%%%%%%%%%%%%%%%%%%%%%%%%%%%%%%%%%%%%%%%%%%%%%%%%%%%%%%%%%%%%

\newcommand{\be}{\begin{equation}}
\newcommand{\ee}{\end{equation}}
\newcommand{\ben}{\begin{eqnarray}}
\newcommand{\een}{\end{eqnarray}}
\newcommand{\nn}{\nonumber \\}
\newcommand{\ii}{\'{\i}}
\newcommand{\pp}{\prime}
\newcommand{\tr}{{\mathrm{Tr}}}
\newcommand{\nd}{{\noindent}}
\newcommand{\grad}{\hspace{-2mm}$\phantom{a}^{\circ}$}

\title{Classical thermodynamics from quasi-probabilities}

\author{\footnotesize F.~Pennini$^{1,2}$\footnote{Corresponding author: flavia.pennini@gmail.com}}

% Affiliations / Addresses
\address{$^{1}$Facultad de Ciencias Exactas y Naturales,
Universidad Nacional de La Pampa, Peru 151, 6300 Santa Rosa,
La Pampa, Argentina \\$^{2}$Departamento de F\'{\i}sica, Universidad Cat\'olica del Norte, Av.~Angamos~0610, Antofagasta, Chile\\
}

\author{\footnotesize  A.~Plastino$^{3}$}

\address{$^{3}$Instituto de F\'{\i}sica La Plata--CCT-CONICET, Universidad Nacional de La Plata, C.C.~727, 1900,
La Plata, Argentina \\
}

\author{ M.C. Rocca$^{3}$}

\address{$^{3}$Instituto de F\'{\i}sica La Plata--CCT-CONICET, Universidad Nacional de La Plata, C.C.~727, 1900,
La Plata, Argentina \\
}

\maketitle

\begin{history}
\received{(Day Month Year)}
\revised{(Day Month Year)}
\end{history}

\begin{abstract}

The basic idea of a microscopic understanding of Thermodynamics is
to derive its main features from a microscopic probability
distribution. In such a vein, we investigate the thermal
statistics of quasi-probabilities's semi-classical analogs in
phase space for the important case of quadratic Hamiltonians,
focusing attention in the three more important instances, i.e.,
those of Wigner, $P$-, and Husimi distributions. Introduction of
an effective temperature permits one to obtain a unified
thermodynamic description that encompasses and unifies the three
different quasi-probability distributions. This unified
description turns out to be classical.
\end{abstract} % end of abstract

%\pacs{03.65.Sq, 05.30.-d,  03.67.-a}

% {03.65.Sq}{Semiclassical theories and applications}   \and
%      {05.30.-d}{Quantum statistical mechanics} \and {03.67.-a}{Quantum information}

 %%%%%%%%%%%%%%%%%%%%%%%%%%%%%%%%%%%%%%%%%%%%%%%%%%%%%%%%%%%%%%%
\section{Introduction}
 %%%%%%%%%%%%%%%%%%%%%%%%%%%%%%%%%%%%%%%%%%%%%%%%%%%%%%%%%%%%%%%

A quasi-probability (QP) distribution is a mathematical
construction resembling  a probability distribution. It   does not
necessarily fulfill, though, some of the Kolmogorov axioms for
probabilities~\cite{plato}.
 QPs display general features of
ordinary probabilities and  yield expectation values with respect
to the weights of the distribution. However, they disobey the
third probability postulate \cite{plato}, in the sense that
regions integrated under them do not represent probabilities of
mutually exclusive states. Some quasi-probability distributions
exhibit zones of negative probability density. QPs often arise in
trying to study  quantum mechanics a phase space representation.
This is frequently done  in quantum optics, time-frequency
analysis, etc.

The dynamics of a quantum system is determined by a master
equation. We speak of an equation of motion for the density
operator ($\hat{\rho}$),  defined with respect to a complete
orthonormal basis.  One can show  that the density can always be
written in a diagonal manner,   with respect to an overcomplete
basis \cite{sudar}. If this basis is that of coherent states
$|\alpha\rangle$ one has~\cite{sudar,glauber}

\be  \label{UNO} \hat{\rho}= \int \frac{d^2\alpha}{\pi}\,
P(\alpha,\alpha^*) \, \vert \alpha \rangle\langle\alpha\vert, \ee
 where  $\mathrm{d}^2
\alpha/\pi=\mathrm{d}x\mathrm{d}p/2\pi\hbar$, with $x$ and $p$
variables of the phase space. The system evolves as prescribed by
the evolution of the quasi-probability distribution function.
Coherent states,   eigenstates of the annihilation operator
$\hat{a}$, serve as the overcomplete basis in such a
build-up~\cite{sudar,glauber}.

There exists a family of different representations, each connected
to a different ordering of the creation and destruction operators
$\hat{a}$ and $\hat{a}^\dagger$. Historically, the first of these
is the Wigner quasi-probability distribution $W$ \cite{wigner},
related to symmetric operator ordering. In quantum optics the
particle number operator is naturally expressed in normal order
and, in the pertinent scenario, the associated representation of
the phase space distribution is the Glauber--Sudarshan $P$
one~\cite{glauber}. In addition to $W$ and $P$, one may find many
other quasi-probability distributions emerging in alternative
representations of the phase space distribution \cite{smooth}. A
quite  popular representation is the Husimi $Q$
one~\cite{husimi,mizrahi1,mizrahi2,mizrahi3}, used when operators
are cast in anti-normal order.

In this paper we wish to apply {\it semiclassical information
theory tools} associated to  these $P$, $Q$, and $W$
representations (for quadratic Hamiltonians) {\it in order to
describe the concomitant thermal semiclassical features}.  We
specialize things to the three $f-$functions associated to a
Harmonic Oscillator~(HO) of angular frequency $\omega$. In such a
scenario the three functions --that we name for sake of
convenience $f_P$, $f_Q$, and $f_W$-- are simple Gaussians and the
treatment becomes entirely analytical, a very convenient feature.
The HO is a really important system that yields insights usually
having a wide impact. Thus, the HO constitutes much more than a
mere simple example. Nowadays, it is of particular interest for
the dynamics of bosonic or fermionic atoms contained in magnetic
traps~\cite{a6,a7,a8} as well as for any system that exhibits an
equidistant level spacing in the vicinity of the ground state,
like nuclei or Luttinger liquids.

  In this paper we
are interested in thermal states regarded as semi-classical
distributions in phase space --analogs of the quantum
quasi-probabilistic distributions. We will try to ascertain what
physical Thermodynamics' features are they able to describe at the
semi-classical level. These distributions
are~\cite{librazo,bookQOptics}

 \ben
f_P=\gamma_P\, e^{-\gamma_P|\alpha|^2},\,\,\,\gamma_P&=&e^{\beta\hbar\omega}-1\,\, \,\,  (P-\textrm{function}),\label{0gama}\\\cr
f_Q=\gamma_Q\, e^{-\gamma_Q|\alpha|^2},\,\,\,\gamma_Q&=&1-e^{-\beta\hbar\omega}\,\,   \,\,(Q-\textrm{function}),\label{1gama}\\\cr
f_W=\gamma_W\, e^{-\gamma_W|\alpha|^2},\,\,\,\gamma_W&=&2\tanh(\beta\hbar\omega/2)\,\,  \,\,  (W-\textrm{function}),\label{3gamas}
\een
 with $\beta=1/k_B T$,  $k_B$ the Boltzmann constant, and $T$ the temperature.
 As stated above, these distributions will be used in the next section as
 semiclassical statistical weight functions.  Since ours is NOT a quantum approach, the ordering
 of the HO-creation and destruction operators $a$ and  $a^{\dagger}$ plays no role whatsoever in our manipulations below.

This paper is organized as follows: section \ref{quantifiers}
refers to  information quantifiers,  in a phase space
representation, for Gaussian distributions. In Section
\ref{Thermo} we focus attention upon thermodynamic relations and
we express them in terms of an effective temperature. Finally,
some conclusions are drawn in Section~\ref{conclu}.

\section{Semi-classical information quantifiers}
\label{quantifiers}

\begin{itemize}

\item   The first step in our development is to calculate  entropic
quantifiers for these Gaussian distributions. In order to simplify
the notation we will consider a general normalized gaussian
distribution in phase space

\be f(\alpha)=\gamma\,e^{-\gamma |\alpha|^2},\label{gaus} \ee
whose normalized  variance is $1/\gamma$ and $\gamma$ taking values $\gamma_P$, $\gamma_Q$
and $\gamma_W$ given by Eqs. (\ref{0gama}), (\ref{1gama}), and (\ref{3gamas}), respectively.

The logarithmic Boltzmann's  information measure for the the
probability distribution  (\ref{gaus}) is

\be S=-\int\,\frac{\mathrm{d}^2 \alpha}{\pi}\,f(\alpha)\,\ln
f(\alpha)=1-\ln \gamma,\label{Shannon} \ee so that it acquires the
particular values

\ben S_P&=&1-\ln\left(e^{\beta\hbar\omega}-1\right),\\\cr
S_Q&=&1-\ln\left(1-e^{-\beta\hbar\omega}\right), \\\cr
S_W&=&1-\ln\left(2 \tanh(\beta\hbar\omega/2)\right), \een for,
respectively, the distributions $f_P$, $f_Q$, and $f_W$.

\item   Next, we focus attention on the information quantifier known as
Fisher's information measure. We  specialize it for families of
shift-invariant distributions, that do not change shape under
translations. One has \cite{bernie,hall} \be I=\,\int
dx\,f(x)\,\left(\frac{\partial \ln f(x)}{\partial x}\right)^2, \ee
and, in phase space, it adopts the appearance \cite{Entropy2014}

\be I=\frac14\,\int
\frac{\mathrm{d}^2\alpha}{\pi}\,f(\alpha)\,\left(\frac{\partial
\ln f(\alpha)}{\partial |\alpha|}\right)^2, \ee such that
considering $f(\alpha)$ given by Eq. (\ref{gaus}) we get
$I=\gamma$,
 whose
specific values are $\gamma_P$, $\gamma_Q$, $\gamma_W$ for the
three functions $f_P$, $f_Q$, and $f_W$, respectively.

\item   The statistical complexity
 is a functional $C[P]$ that can primarily be viewed as a
quantity that characterizes the probability distribution ${P}$.
 It quantifies not only randomness but also the
presence of correlational structures \cite{LMC}. The opposite
extremes of perfect order and maximal randomness possess no
structure to speak of. In between these two special instances, a
wide range of possible degrees of physical structure exist,
degrees that should be reflected in the features of the underlying
probability distribution.  The statistical complexity $C$,
according to L\'{o}pez-Ruiz, Mancini, and Calvet \cite{LMC},  is a
suitable product of two quantifiers, such that $C$ becomes minimal
at the extreme situations of perfect order or total randomness.
Instead of using the prescription of \cite{LMC}, but without
violating its spirit,
 we will take one of these two  quantifiers to be Fisher's measure and  the other an entropic form, since it
 is well known that the two behave in opposite manner~\cite{frieden}. Thus:

\be C=S I= \gamma\, (1-\ln \gamma), \ee that vanishes for perfect
order or total randomness. For each particular case, we explicitly
have

\ben
C_P&=&\left(e^{\beta\hbar\omega}-1\right)\left[1-\ln\left(e^{\beta\hbar\omega}-1\right)\right],\\\cr
C_Q&=&\left(1-e^{-\beta\hbar\omega}\right)\left[1-\ln\left(1-e^{-\beta\hbar\omega}\right)\right],
\\\cr C_W&=&\left(2
\tanh(\beta\hbar\omega/2)\right)\left[1-\ln\left(2
\tanh(\beta\hbar\omega/2)\right)\right], \een for, respectively,
the distributions $f_P$, $f_Q$, and $f_W$. The maximum of the
statistical complexity occurs when $\gamma=1$ and, the associated
temperature values are
\begin{displaymath}
 \left\{ \begin{array}{ll}
e^{\beta\hbar\omega}-1=1\Rightarrow T=\hbar\omega/k_B\ln2 &\,\,\,\, \textrm{for the $f_P-$function},\\\\
1-e^{-\beta\hbar\omega} =1\Rightarrow T=0& \,\,\,\,\textrm{for the $f_Q-$function},\\\\
2\tanh(\beta\hbar\omega/2)=1 \Rightarrow
T=\hbar\omega/2k_B\arctan(1/2)&\,\,\,\, \textrm{for the
$f_W-$function}.
\end{array} \right.
\end{displaymath}

\end{itemize}

%%%%%%%%%%%%%%%%%%%%%%%%%%%%%%%%%%%%%%%%%%%%%%%%%%%%%%%%%%%%%%%%%%
\section{Thermodynamic relations}
%%%%%%%%%%%%%%%%%%%%%%%%%%%%%%%%%%%%%%%%%%%%%%%%%%%%%%%%%%%%%
\label{Thermo}

 We start this section considering the semi-classical Hamiltonian of the harmonic oscillator that reads

\be \mathcal{H}(x,p)= \hbar \omega |\alpha|^2=\hbar\omega \left(\frac{x^2}{4\sigma_x^2}+\frac{p^2}{\sigma_p^2}\right), \label{H0}
  \ee where $x$ and $p$ are  phase space variables and
 $\sigma_x^2=\hbar/2m\omega$ and $\sigma_p^2=\hbar
m\omega/2$~\cite{pathria}. Let us further define the  semiclassical expectation
value of the function $\mathcal{A}(x,p)$ as

\be \langle \mathcal{A}\rangle_f=\int\frac{\mathrm{d}^2
\alpha}{\pi}\,f(\alpha)\,\mathcal{A}(x,p), \label{mvalues}\ee
indicating that $f(\alpha)$ is the statistical weight function.
Thus, the mean energy  of the hamiltonian $\mathcal{ H}(x,p)$ is
written in the fashion

\be U^*=\hbar\omega\,\int\,\frac{\mathrm{d}^2
\alpha}{\pi}\,f(\alpha)\,|\alpha|^2=\frac{\hbar\omega}{\gamma},\label{em}
\ee where   $\gamma$ takes the respective values $\gamma_P$,
$\gamma_Q$, and $\gamma_W$ explained in Introduction.
Additionally, the thermodynamic entropy $S^{'}$  is

\be S^{'}=k_B(1-\ln \gamma),\label{ent_term} \ee where we have
added the Boltzmann constant $k_B$. The mean energy can be viewed
as a function of the thermodynamic entropy $S^{'},$ in the
following fashion. Combining (\ref{em}) with the thermodynamic
entropy (\ref{ent_term})~we~get the associated, fundamental
equation  $U^*=U^*(S^{'})$

\be U^*(S^{'})=\hbar\omega\, e^{S^{'}/k_B-1}, \ee and \be
\gamma=e^{1-S^{'}/k_B}. \ee Thus, the differential of
$U^*$~becomes

\be \mathrm{d}U^*=\left(\frac{\partial U^*}{\partial
S^{'}}\right)_{V}\,\mathrm{d}S^{'}, \ee where we have considered the
volume $V$ to be constant.  Thus, after effecting the pertinent
replacements  we find \be \mathrm{d}U^*=\frac{\hbar\omega}{k_B
\gamma}\,\mathrm{d}S^{'}, \ee which suggests introducing an effective
temperature  $T_{eff}$. {\it Using $T_{eff}$ we obtain a unified
picture that encompasses the three distributions $f_P$, $f_Q$, and
$f_W$, in a single thermodynamic description}. We have

\be T_{eff}=\left(\frac{\partial U^*}{\partial
S^{'}}\right)_V=\frac{\hbar\omega}{k_B \gamma},\label{0teff} \ee such
that

\be \mathrm{d}U^*=T_{eff}\,\mathrm{d}S^{'}.\label{teff} \ee

Note that in the three instances,  $T_{eff}= \infty$ for
$T=\infty.$ However, if $T=0$,  $T_{eff}=0 $ only in the $f_P$-case.
It  equals $1/2$ in the Wigner instance and equals $1$ in the
Husimi case, as depicted in the accompanying figure.

\begin{figure}[h]
\begin{center}
\includegraphics[scale=0.6,angle=0]{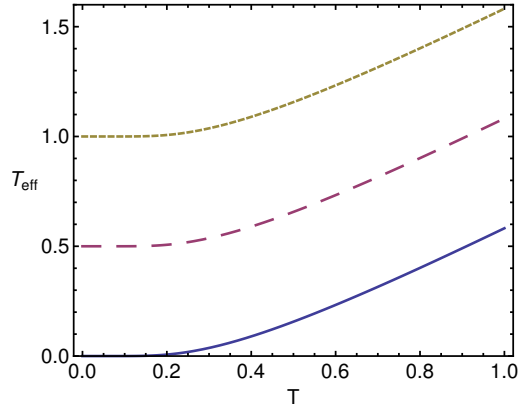}
\vspace{-0.2cm} \caption{Effective temperature $T_{eff}$ versus
temperature $T$ in $(\hbar \omega/k_B)-$units. }
\end{center}
\end{figure}
From (\ref{em}) and (\ref{teff}) we can rewrite the mean energy in
terms of effective temperature.

\be U^*=k_B\, T_{eff}, \label{Ueff}\ee that corresponds to the
classical mean energy of a harmonic oscillator of temperature
$T_{eff}$, with $k_B\, T_{eff}/2$ contributions for each of the
two pertinent degrees of freedom. Similarly, the thermodynamic
entropy is recast as \be \frac{S^{'}}{k_B}=1+\ln\left(\frac{k_B\,
T_{eff}}{\hbar\omega}\right), \ee and the Helmholtz free energy is
given by

\be A^*=U^*-T_{eff}\,
S^{'}=k_BT_{eff}\,\ln\left(\frac{\hbar\omega}{k_B\,
T_{eff}}\right).\label{free} \ee The effective specific heat is
defined as

\be C_V^{*}=\left(\frac{\partial U^*}{\partial T_{eff}}\right)_V,
\ee that using (\ref{Ueff})  becomes

\be C_V^{*}=k_B, \ee which is precisely the specific heat for the
classical harmonic oscillator which is independent of the
temperature. This becomes the  Dulong and Petit's rule at the
classical limit. In view of (\ref{teff}) and (\ref{free}) the
analog partition function $Z^*$ is given~by

\be Z^*=\frac{1}{\gamma},\label{part} \ee and,
according to Eqs. (\ref{Shannon}), (\ref{em}), and (\ref{part}) we
find

\be S^{'}=\ln Z^*+\beta^{*} U^*, \ee with

\be \beta^{*}=\frac{1}{k_B T_{eff}}=\frac{\gamma}{\hbar\omega}.
\ee Thus, {\sf one reobtains all the thermal results pertaining to
a classical HO at the temperature $T_{eff}$. Note that the whole
thermal description becomes now of a classical character. All the
quantum effects pertaining to  the probability distributions are
contained in the relationship (24) between $T_{eff}$ and $T$}.

\vskip 4mm

It is interesting to look at the statistical complexity $C$ in
order to see at what effective temperature the possible
correlational structures carried by our probabilistic
distributions are stronger. Expressed in terms of $T_{eff}$, $C$
becomes

\be C^{'}=I S^{'}
=\frac{\hbar\omega}{T_{eff}}\,\left[1+\ln\left(\frac{k_B\,
T_{eff}}{\hbar\omega}\right)\right]. \ee Keeping in mind
$T_{eff}$'s definition, it is easy to see that the maximum for the complexity $C^{'}$
is attained when $ T_{eff}=\hbar\omega/k_B.$ This implies, according
to Eq. (\ref{0teff}) that the maximum of the Fisher measure es $I_{max}=1$. At the complexity-peak,
thermodynamic quantities take the values

\ben U^*_{max}&=&\hbar\omega,\\\cr S^{'}_{max}&=&k_B,\\\cr
 C^{'}_{max}&=&k_B, \een a remarkable simplicity!

%%\newpage
\section{Conclusions}
%%%%%%%%%%%%%%%%%%%%%%%%%%%%%%%%%%%%%%%%%%%%%%%%%%%%%%%%%%%%%%%%
\label{conclu}

We have investigated here the thermal statistics of
quasi-probabilities-analogs $f(\alpha)$ in phase space for the
important case of quadratic Hamiltonians, focusing attention on
the three more important instances, i.e., those of Wigner, $P$-,
and Husimi distributions.

\begin{itemize}

\item Introduction of an effective temperature permits one to
obtain a unified thermodynamic description that encompasses the
three different quasi-probability distributions. This unified
description turns out to be classical.

\item The above entails that all possible ``quasi-quantum" effects have to be contained in the
relationship between $T_{eff}$ and $T$. Note, for instance, that
the minimal energy  is not zero ({\it one of these effects}) but,
in the $f_W-$case, $T^W_{eff}= \hbar\omega/2k_B$, implying a
minimum energy $k_B T^W_{eff}= \hbar \omega/2$. Additionally, the
Husimi-$T^Q_{eff}= \hbar\omega/k_B$ reflects the well known fact
that the Husimi distribution ``smoothes" the Wigner one over a
phase-space area $=\hbar$.

\item The basic idea of a microscopic understanding of thermodynamics is
to derive its main features from a microscopic probability
distribution. We have done just this using as distributions
quasi-probability ones.

\end{itemize}

\end{document}